\documentclass[conference,compsoc]{IEEEtran}
\IEEEoverridecommandlockouts

\ifCLASSOPTIONcompsoc
  \usepackage[nocompress]{cite}
\else
  \usepackage{cite}
\fi

\ifCLASSINFOpdf
\else
\fi
\usepackage{algorithmic}
\usepackage{graphicx}
\usepackage{textcomp}
\usepackage{xcolor}
\usepackage{tcolorbox}
\usepackage{braket}
\usepackage[hyphens]{url}
\usepackage{cite}
\usepackage{enumitem}
\usepackage[bookmarks=true,breaklinks=true,letterpaper=true,colorlinks,citecolor=blue,linkcolor=blue,urlcolor=blue]{hyperref}
\def\BibTeX{{\rm B\kern-.05em{\sc i\kern-.025em b}\kern-.08em
    T\kern-.1667em\lower.7ex\hbox{E}\kern-.125emX}}

\usepackage{multirow}
\usepackage{multicol}

\usepackage{tikz}
\newcommand*\circled[1]{\tikz[baseline=(char.base)]{
            \node[shape=circle,draw,inner sep=1pt,fill=white, text=black] (char) {#1};}}
\newcommand{\ignore}[1]{}
\newcommand{\review}[1]{\color{red}#1 \color{black} }
\newcommand{\framework}{COMPASS}
\newcommand{\frameworkext}{E-COMPASS}

\usepackage{array}
\begin{document}

\title{COMPASS:  Compiler Pass Selection \\For Improving Fidelity Of NISQ Applications}

\author{\IEEEauthorblockN{Siddharth Dangwal\IEEEauthorrefmark{1}, Gokul Subramanian Ravi \IEEEauthorrefmark{2}, Lennart Maximilian Seifert\IEEEauthorrefmark{1}, \\ Poulami Das\IEEEauthorrefmark{3},  James Sud\IEEEauthorrefmark{1} and Frederic T. Chong\IEEEauthorrefmark{1}} 
\IEEEauthorblockA{\IEEEauthorrefmark{1}Department of Computer Science, University of Chicago} \IEEEauthorblockA{\IEEEauthorrefmark{2}Department of Computer Science and Engineering, University of Michigan} \IEEEauthorblockA{\IEEEauthorrefmark{3}Department of Electrical and Computer Engineering, The University of Texas at Austin}
}

\maketitle

\IEEEpubid{\makebox[\columnwidth]{979-8-3315-4127-9/24/\$31.00~\copyright~2024 IEEE \hfill}
\hspace{\columnsep}\makebox[\columnwidth]{ }}

\begin{abstract}
Noisy qubit devices limit the fidelity of programs executed on near-term or \textit{Noisy Intermediate Scale Quantum (NISQ)} systems. The fidelity of NISQ applications can be improved by using various optimizations during program compilation (or \textit{transpilation}). These optimizations or \textit{passes} are designed to minimize circuit depth (or program duration), steer more computations on devices with lowest error rates, and reduce the communication overheads involved in performing two-qubit operations between non-adjacent qubits. Additionally, standalone optimizations have been proposed to reduce the impact of crosstalk, measurement, idling, and correlated errors. However, our experiments using real IBM quantum hardware show that using all optimizations simultaneously often leads to sub-optimal performance and the highest improvement in application fidelity is obtained when only a subset of passes are used. 
Unfortunately, identifying the optimal pass combination is non-trivial as it depends on the application and device specific properties.\\ 
In this paper, we propose {\em \framework{}}, an automated software framework for optimal {\em \underline{Com}piler \underline{Pas}s \underline{S}election} for quantum programs.  {\em \framework{}} uses dummy circuits that resemble a given program but is composed of only Clifford gates and thus, can be efficiently simulated classically to obtain its correct output. The optimal pass set for the dummy circuit is identified by evaluating the efficacy of different pass combinations and this set is then used to compile the given program.  Our experiments using real IBMQ machines show that {\em \framework{}}  improves the application fidelity by 4.3x on average and by up-to 248.8x compared to the baseline. However, the complexity of this search scales exponential in the number of compiler steps. To overcome this drawback, we propose  {\em Efficient \framework{}} ({\em \frameworkext{}}) that leverages a divide-and-conquer approach to split the passes into sub-groups and exhaustively searching within each sub-group. Our evaluations show that \frameworkext{} improves fidelity by 3.0x on average and by up-to 257.1x compared to the baseline while reducing {\em \framework{}} overheads by 200x and up to 327x.
\end{abstract}

\section{\textbf{Introduction}}\label{sec:introduction}

Today, quantum computers with only a few dozen qubits can outperform the state-of-the-art supercomputers for some specific sampling tasks~\cite{kim2023evidence, arute2019quantum}. These machines are projected to surpass a few thousand qubits by the end of 2025~\cite{ibm} and are expected to power real-world applications soon. 
However, noisy quantum hardware presents a significant challenge in achieving this goal. The average error rate of a quantum operation ranges between 0.1-1\% on existing systems~\cite{device_specs}. Consequently, program execution on these systems encounters errors limiting the fidelity of applications. Unfortunately, near-term or \textit{Noisy Intermediate Scale Quantum (NISQ)} systems cannot achieve complete fault-tolerance.   Instead, they run programs in the presence of errors and yet promise to accelerate some domain-specific problems~\cite{adaptvqe, farhi2014quantum}. To leverage NISQ systems to solve real-world problems, the application fidelity must be improved by mitigating the impact of hardware errors.

\begin{figure*}
    \centering
    \includegraphics[width=\textwidth]{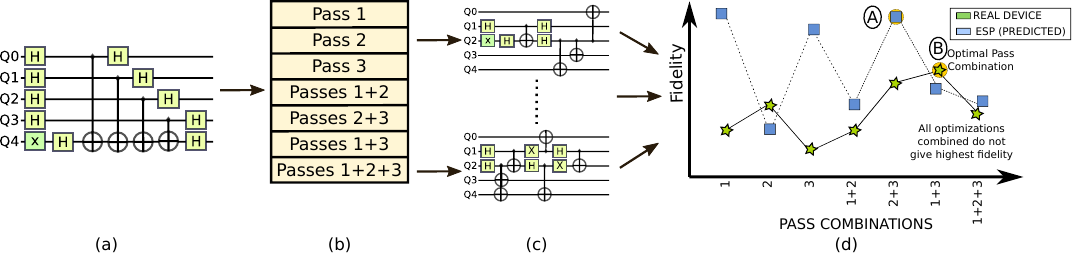} \vspace{-0.25in}
    \caption{(a) A quantum circuit- it must be before execution on hardware (b) All possible pass combinations that can be used to compile (assume three passes - Passes 1, 2, and 3) (c) Each combination creates a different executable. Each schedule is impacted by hardware errors differently. (d) The combination that corresponds to the highest fidelity (marked B) may be a subset of all ``optimizations". It is hard to predict this optimal set using analytical tools like estimating the Expected Success Probability (ESP) of executing a compiled schedule. The performance of the combination predicted by ESP (marked A) is sub-optimal. }\vspace{-0.15in}
    \label{fig:intro}\vspace{-0.08in}
\end{figure*}

The impact of hardware errors can be reduced by generating highly optimized machine code during program \textit{compilation} or \textit{transpilation}. Compilation is a multi-step process that maps program qubits onto the physical qubits of a given device, schedules instructions by maintaining the program's data dependencies, and translates high-level operations into low-level machine-compliant native gates. Additionally, compilers introduce \textit{routing} or \textit{SWAP} operations to relocate non-adjacent qubits to 
neighboring locations so that two-qubit CNOT gates can be performed between them. These functionalities are completed in a step-by-step fashion, where at each step, a program is acted upon by an optimization (or \textit{pass}), finally resulting in an executable that can be run on a quantum device with greater fidelity compared to an unoptimized version.

There exists several compiler passes that reduce the number of computations, lower program depth, increase operational concurrency, and enable noise-aware computations by taking into account the hardware error characteristics~\cite{murali2019noise, li2019tackling, pokharel2018demonstration, viola1999dynamical}.  Naively, these passes can be viewed as utilities that uniformly improve the fidelity of every program. In that case, the most intuitive choice would be to \textit{apply all of them} simultaneously to compile a  program. However, our experiments on real systems from IBM show that this approach does not maximize application fidelity. Rather, we routinely achieve the highest fidelity when we apply only a subset of the compiler passes. For example, the fidelity of the multi-control cnot (cnxh7) benchmark is sub-optimal when three optimization passes are combined, whereas the highest fidelity is obtained when only two are used of these passes are used, as shown in Figure~\ref{fig:intro}.



The performance gap between the optimal set of passes and all combined arises from the imperfections in the heuristics used to design passes. Most passes are developed standalone by choosing a heuristic that only optimizes for a single objective. When the heuristics of multiple passes are combined, they interfere with each other, often destructively, lowering the performance. For example, SABRE~\cite{li2019tackling} is a routing algorithm that minimizes SWAPs and maximizes CNOT concurrency to reduce circuit depth. However, its heuristics do not account for operational crosstalk and may schedule two CNOTs in parallel that has high effective error-rate due to crosstalk. An alternate optimization~\cite{murali2020software} reduces CNOT crosstalk but has limited efficacy as it only avoids concurrent CNOTs on two neighboring (local) links if they have high crosstalk. Consequently, using~\cite{murali2020software} cannot assess the impact of crosstalk in the routing paths selected by SABRE to relocate two distant qubits. \textit{Thus, it becomes crucial to identify the optimal subset of passes that maximizes application fidelity when combined.}

However, identifying the optimal set of compiler passes is challenging due to several reasons. \textit{First}, it is specific to an application and device. \textit{Second}, it varies over time as the systems characteristics change due to device drifts and machine calibrations. \textit{Third}, selection based on analytical fidelity estimation metrics such as \textit{Expected Success Probability (ESP)} used by existing compilers~\cite{tannu2019not, tannu2019ensemble, nation2022suppressing} does not capture the complex trends associated with the interference of passes because it only computes the probability of successfully executing a schedule assuming all operations remain error-free and no qubits decohere. Consequently, the predicted fidelity is often far from the true fidelity observed on real systems. For example, Figure~\ref{fig:intro}(d) shows the actual fidelity and the ESP for a multi-control cnot benchmark on IBMQ-Toronto for different combinations of three representative passes. We observe that the combination selected by ESP (shown as \circled{A}) is different from the optimal combination (shown as \circled{B}). \textit{Lastly}, we cannot use device characterization and noisy simulations because they do not scale. \textit{Ideally, we want to accurately identify the optimal set of compiler passes specific to an application and a given device in a scalable manner.}


In this paper, we propose {\em \framework{}}, a software framework that selects the optimal set of passes to compile a quantum circuit for attaining maximum improvements in fidelity. Hypothetically, if the solution of a program was known, we could identify the optimal pass set by testing the efficacy of different pass combinations on a real system. However, this is not possible because the program output is unknown. \framework{} addresses this limitation by leveraging two key insights. \textit{First}, most pass heuristics make decisions based on the positions of CNOT gates in the circuit. \textit{Second}, CNOT gates are Clifford operations and circuits with only Clifford gates can be efficiently simulated on classical machines to identify its correct output~\cite{gottesman1998heisenberg}. 
\framework{} translates a given program (which consists of both Clifford and non-Clifford gates) into a \textit{dummy circuit} with only Clifford gates. \footnote{Clifford dummies have been previously used to design some specific  passes~\cite{das2021adapt, das2023imitation}. However, their proposed usage in \framework{} is unique and orthogonal to these works. We discuss this in Related Work Section~\ref{sec:discussion}.} As CNOT is a Clifford gate, the position of all CNOTs in the program and the Clifford dummy is same. Both circuits suffer from similar noise effects as CNOTs are the dominant sources of errors in quantum programs. \framework{} evaluates the efficacy of all possible pass combinations and identifies the one that maximizes the dummy circuit fidelity. \framework{} uses this subset of passes to compile the given program.

\ignore{
\footnote{Clifford dummies have been previously used to design some specific individual passes~\cite{}. However, their proposed usage in COMPASS is unique and orthogonal to these prior works. We discuss them in details in Section~\ref{}.   
}
}

\begin{figure*}[ht]
    \centering
    \includegraphics[width=2\columnwidth]{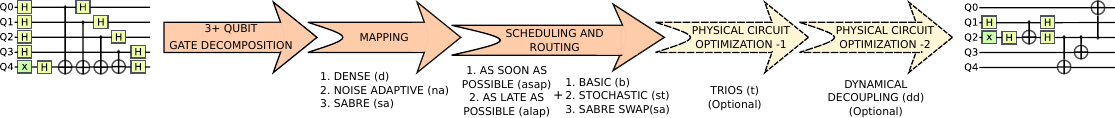}\vspace{-0.05in}
    \caption{The IBM-Qiskit transpiler pipeline. The details of each pass are discussed in Section\ref{subsec:transpilation}. The transpiler takes an abstract quantum circuit as input and gives an optimized physical quantum circuit as output. The first three steps of the pipeline are compulsory. The physical circuit optimizations are optional.}
    \label{fig:transpilation_pipeline}\vspace{-0.2in}
\end{figure*}

\framework{} exhaustively evaluates the impact of all possible pass combinations on the fidelity of the dummy circuit. This limits its scalability as the search space grows exponentially in the number of transpiler steps. To reduce this search space, we propose {\em Efficient \framework{}} or {\em \frameworkext{}} that accomplishes the search using a divide-and-conquer approach. \frameworkext{} splits the total set of passes into smaller \textit{groups} and exhaustively searches for the best combination within each group. The optimal set of passes for the program is then obtained by combining the best combinations from each group. Using this approach, \frameworkext{} reduces the complexity to \textit{linear} in the number of transpiler steps. 
Also, \frameworkext{} leverages the insight that the number of shots (or trials) required to evaluate the efficacy of each pass combination on the dummy circuit can be significantly reduced because we only need a ``good enough" fidelity estimation and the relative performance of pass combinations against each other.

Our evaluations on 27-qubit IBM quantum computers show that \framework{} increases program fidelity by 4.3x on average and by up to 248.8x in the best-case over the ``best" compilation scheme used by IBM's Qiskit compiler. \frameworkext{} improves fidelity by 3.0x on average and by up to 257.1x while requiring 3.26x fewer circuit evaluations than \frameworkext{}. It reduces the overheads further by 61x on average and upto 100x via shot reduction.  

To that end, this paper makes the following contributions:
\begin{enumerate}[leftmargin=0cm,itemindent=.5cm,labelwidth=\itemindent,labelsep=0cm,align=left, itemsep=0.1 cm, listparindent=0.4cm]
    \item We show that applying all available compiler passes does not maximize NISQ program fidelity and the highest fidelity is obtained when only a subset of passes is used. 
    \item We show that the optimal set of passes is specific to a program as well as quantum hardware and varies over time. 
    \item We propose {\em \framework{}}, a software framework that identifies the optimal pass combination for maximizing the fidelity of any arbitrary circuit by using Clifford dummy circuits.
    \item We propose {\em Efficient \framework{} (\frameworkext{})} that uses a divide-and-conquer search algorithm and approximate fidelity estimation for greater scalability. 
\end{enumerate}


\section{\textbf{Background}}\label{sec:background}

\subsection{\textbf{Quantum Errors}}\label{subsec:qubits_and_gates}

A qubit is the fundamental unit of information in quantum computing. Quantum computers leverage the properties of \textit{superposition} and \textit{entanglement} to enable efficient algorithms. In most programs, the qubits are initialized to a superposition state followed by the application of \textit{quantum gates} to manipulate their probability amplitudes such that the probabilities associated with the correct outcome(s) are boosted. Finally, the desired outcomes are obtained by \textit{measuring} the qubits. However, on real systems, programs frequently produce incorrect outcomes because 
qubit devices are noisy. For example, the average error-rates of a two-qubit CNOT gate and measurement operation is around $1.2\%$ and $2.6\%$~\cite{IBMQE} respectively on a 5-qubit IBMQ-Lima machine. Qubit devices also accumulate errors when left idle and cannot retain information beyond few tens of micro-seconds on systems from IBM or Google. Moreover, error rates vary across devices and over time~\cite{patel2020experimental}. 

\subsection{\textbf{Primer on Quantum Transpilation}}\label{subsec:transpilation}
Quantum programs are written at a high-level to facilitate seamless software development, whereas quantum hardware supports only a limited number of low-level basis gates. \textit{Compilation} or \textit{transpilation} is a necessary step that transforms a high-level program into a functionally equivalent low-level executable compliant with the target machine. Similar to classical compilation, quantum compilers too perform their tasks in a step-by-step fashion. These steps are represented as \textit{passes}.  
The compiler takes a quantum circuit as input and applies passes in a specific order to produce an executable that respects the constraints of a specific device (for instance qubit connectivity and native gate set). Figure~\ref{fig:transpilation_pipeline} shows an overview of compilation and some of the steps are described next. 

\begin{itemize}[leftmargin=0cm,itemindent=.5cm,labelwidth=\itemindent,labelsep=0cm,align=left, itemsep=0.1 cm, listparindent=0.4cm]
    
    \item \textbf{3+ Qubit Gate Decomposition}: Typically superconducting quantum hardware can support only single- and two-qubit gates. Gates that act on more than two qubits must be decomposed as a combination of single and two-qubit gates.
    
    \item \textbf{Mapping}: Program qubits must be mapped to the physical qubits on the device. The fidelity of a circuit depends on the mapping as qubits exhibit variability in errors. \textit{Good mappings} reduce the usage of devices with high error-rates.

     \item \textbf{Scheduling and Routing}: The compiler schedules instructions by maintaining a program's data dependencies. Routing is a part of scheduling CNOTs involving non-adjacent qubits. As quantum systems often have limited connectivity, two non-adjacent qubits must be routed next to each other by using SWAP gates to perform CNOTs between them. A SWAP is a sequence of three CNOTs and further increases the vulnerability to errors. A \textit{good routing} pass tries to minimize the errors introduced by SWAPS.
    
    \item \textbf{Physical Circuit Optimizations}: These are optimizations we apply to the circuit post mapping and routing to further suppress noise. For example, padding idle spaces with dynamical decoupling pulses helps in reducing idle errors~\cite{pokharel2018demonstration}. 
    
\end{itemize}

\subsection{\textbf{Transpiler Passes for Error Mitigation}}\label{subsec:passes_for_em}

The transpiler's primary objective is to transform a circuit in such a way that device constraints are satisfied. However most passes do more than the bare minimum in this direction. They often make decisions in a way such that the overall fidelity of the circuit is maximized when run on the NISQ device. 

A straightforward way to improve circuit fidelity while mapping is to choose physical qubits ``close" to each other to reduce routing overheads as SWAPs are extremely error-prone. For example, the error rate of a SWAP gate is about $3.5\%$ on IBMQ-Lima. The \textbf{Dense mapper} chooses the most densely connected subgraph which is isomorphic to the quantum circuit connectivity graph. An alternate strategy called \textbf{Noise-Adaptive} mapping accounts for the variable error-rates of CNOT links and measurements to choose better-than-worst case ones \cite{murali2019noise, tannu2019not} in a greedy fashion. \textbf{SABRE} \cite{li2019tackling} is another approach that starts with random mapping and iteratively computes a better one by reducing the SWAP costs~\cite{li2019tackling}. 

The \textbf{As Soon As Possible} (ASAP or \texttt{al}) policy schedules operations as soon as their dependencies are resolved, whereas the \textbf{As Late As Possible} (ALAP or \texttt{al}) scheduling waits to schedule an operation to the point where it is absolutely necessary to execute it to make computational progress.
Routing policies such as \textbf{Stochastic} routing, \textbf{Basic} routing, and \textbf{SABRE} routing specifically aim to minimize SWAPs. Stochastic routing employs a hit-or-miss strategy where random SWAPs are inserted between unconnected qubits and the pass terminates when it finds a configuration that places all interacting qubits next to each other at the right time instances. Since this search does not have a guiding principle, the results can vary substantially across runs. Basic routing finds the shortest route between two non-interacting qubits, routes them along that path, and takes them back to their initial positions.

There also exist policies that break the transpilation flow abstraction to achieve lower gate count. \textbf{Trios} \cite{duckering2021orchestrated} decomposes three-qubit Toffoli gates into single- and two-qubit gates after mapping and routing, and not before as is described in the conventional strategy in Section \ref{subsec:transpilation}. This reduces gate count and circuit depth, and consequently improves circuit fidelity. Another scheme that can be incorporated as part of the transpilation pipeline is padding the routed circuit with dynamical decoupling pulses \cite{das2021adapt, padDD} to reduce idle errors.

\section{\textbf{Challenges In Optimal Pass Set Selection}}\label{sec:motivation}


Every quantum circuit compulsorily needs to be \textit{mapped}, \textit{scheduled}, and \textit{routed} in order to be executed on a device. Additional passes may also be used to further improve the application fidelity such as insertion of dynamical decoupling pulses. The objective of the characterization experiments presented in this Section is to evaluate the efficacy of different pass combination on the application fidelity and the complexity of this study scales exponential in the number of passes. It also requires several circuit executions that are bottle-necked by long cloud latencies. To limit this complexity for the scope of our study, we choose mainly the mandatory Mapping-Scheduling-Routing (M-S-R) combinations. \textit{Note that our final evaluations includes other optional passes as well.}

\begin{figure}[!h]
    \centering
    \includegraphics[width=0.85\columnwidth]{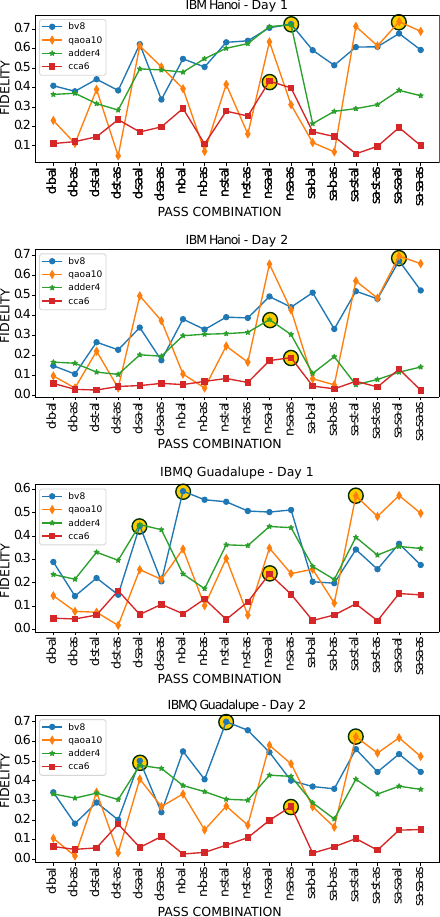}
    \caption{Variation of optimal mapping-scheduling-routing pass set for four circuits on IBM Hanoi and IBMQ Guadalupe on 2 Sep 2022 and 8 Sep 2022. The best pass combination in each subplot is highlighted with a yellow circle.}
    \label{fig:mrs_variation}
\end{figure}


\subsection{\textbf{Optimal Pass Set Varies with Circuit, Device, and Time}}\label{subsec:variation_in_pass_combos}

IBM's Qiskit transpiler offers (1)~three mapping options: dense (\texttt{d}), noise-adaptive (\texttt{na}) and sabre (\texttt{sa}), (2)~two scheduling options: As Late As Possible (\texttt{al}) and As Soon As Possible (\texttt{as}), and (3)~four routing options: basic swap (\texttt{b}), stochastic swap (\texttt{st}), look-ahead swap and sabre swap (\texttt{sa}). Passes like noise-adaptive and sabre are state of the art mappers and routers. We refer the readers to~\cite{IBMTranspiler} for a comprehensive understanding of these passes. To understand the performance of different M-S-R combinations, we take four representative circuits, compile each of them using different M-S-R combinations, run them on multiple IBM machines, and compute their fidelities. Note that sabre routing already includes lookahead heuristics during SWAP selection. We do not use IBM's exhaustive lookahead SWAP owing to its exponential complexity. It fails to produce an executable for most of our benchmarks. Thus, we get a total of 18 M-S-R combinations (3 mappers, 2 schedulers, and 3 routers).

\vspace{0.05in}
\noindent \textbf{Observation-1:} Figure \ref{fig:mrs_variation} shows that the \textit{optimal transpiler pass set} or combination differs across circuits even on the same day and the same machine. This is expected as each circuit has a different structure, is mapped to different links and qubits, and thus encounters different levels of noise. 

\vspace{0.05in}
\noindent \textbf{Observation-2:} For a given circuit, on a given day, the optimal pass combination differs across machines. This is also expected as there is a lot of heterogeneity in superconducting chips and their error profiles differ substantially \cite{IBMQE}.

\vspace{0.05in}
\noindent \textbf{Observation-3:} For a given circuit and device, the optimal pass set varies over time too. This too, is expected because the error characteristics of devices drift over time \cite{patel2020experimental, IBMQE}. 

\vspace{0.05in}
\noindent \textbf{Observation-4:} IBM Qiskit's default  highest optimization level (\texttt{3}) uses Sabre mapping, Sabre routing, and ALAP scheduling (\texttt{sa-sa-al}) and this is often not the best policy.

\begin{figure}[b]
    \centering\vspace{-0.1in}
    \includegraphics[width = 0.8\columnwidth]{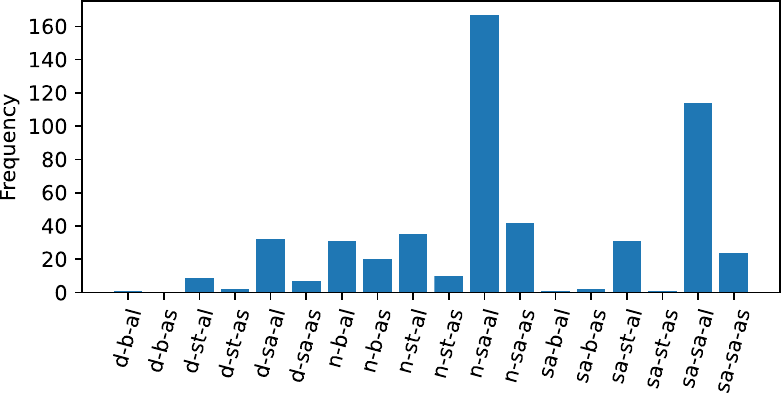}\vspace{-0.1in}
    \caption{Frequency of different optimal M-S-R combinations, when computed over 4 device backends, 13 days and 13 circuits.}
    \label{fig:motivation_frequency}
\end{figure}

\begin{figure*}[!htp]
    \centering
    \includegraphics[width=\textwidth]{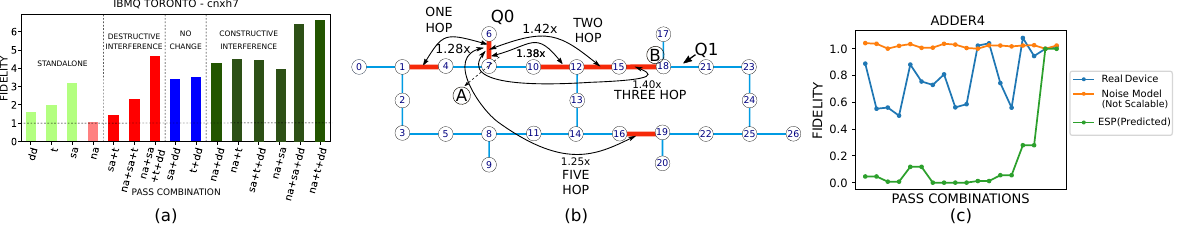}\vspace{-0.05in}
    \caption{(a) Pass Interactions for 7 qubit \texttt{cnxh7} benchmark on IBMQ-Toronto (b) Increase in error rate of the (6,7) CNOT due to crosstalk from other CNOT links in the circuit. We see a substantial increase in error due to crosstalk between CNOT gates that are 3-5 hops away (c) Fidelity trends for device execution, noise model and ESP for the ADDER4 benchmark. The X-axis represents different transpiler pass combinations.}\vspace{-0.05in}
    \label{fig:combined_motivation_figure}
\end{figure*}

\vspace{0.05in}
\noindent \textbf{Observation-5:} The ``best" pass set is not limited to just a small subset of all possible choices. Figure \ref{fig:motivation_frequency} shows the frequency that each M-S-R combination is ``optimal" for thirteen circuits on four machines over thirteen days. While \texttt{na-sa-al} and \texttt{sa-sa-al} (Qiskit's highest \texttt{optimization level 3}) are the most frequent options, being reported optimal $57\%$ of the time, others occur with a non-negligible probability of $43\%$. Thus, a device characterization approach using some benchmarks cannot be used.

\subsection{\textbf{Passes Interact Both Constructively and Destructively}}\label{subsec:pass_interactions}
Compiler passes interact in non-intuitive ways. Naively, we expect two passes that improve fidelity over the baseline to work in synergy to boost it further. Similarly passes that lower circuit fidelity would combine to produce a result that is worse compared to the baseline. A combination of a ``good" and a ``bad" pass might or might not improve fidelity depending on which pass dominates. However, that is often not the case.

Figure \ref{fig:combined_motivation_figure}(a) gives an example of how passes interact for the cnxh7 benchmark. As this is a multi-control CNOT circuit, we also use the Trios (\texttt{t}) pass in addition to the passes described in Section~\ref{subsec:variation_in_pass_combos} because it is particularly suited for such circuits. Additionally, as this circuit has large depth and high idle times, we also use the dynamical decoupling (\texttt{dd}) pass. This enables us to have a fair comparison. 


\vspace{0.05in}
\noindent \textbf{Observation-6:} We observe that noise\_adaptive mapping (\texttt{na}) standalone degrades fidelity (happens when calibration data is not up to date and the device has drifted). However, when combined with (\texttt{t}) and (\texttt{dd}), it produces the highest fidelity (6.5x) across all pass combinations. Compare this with \texttt{sa} and \texttt{t}, which standalone gives 3x and 2x improvements respectively, together they give no fidelity improvement (sa+t $\sim$ 1x). Similarly, when \texttt{sa} is combined with \texttt{na+t+dd} the fidelity reduces despite each improving it standalone. 
Figure \ref{fig:combined_motivation_figure}(a) shows the regions of ``destructive interference", where the net result potentially improves when we take away one of the constituent passes in the set. It also shows all cases of ``constructive interference". This refers to all cases where the combined effect of a set of passes is greater than the sum of the improvements caused by the constituents. We also observe some combinations whose net effect is broadly the same as that of one of its constituents.

\ignore{
\begin{tcolorbox}[width=\columnwidth,colback=blue!5!white, colframe=blue!0!white]
\textbf{Passes that independently improve circuit fidelity can act in conjunction to reduce it. Passes that individually degrade circuit fidelity when combined can potentially improve it.}
\end{tcolorbox}}

\subsection{\textbf{Compiler Passes Rarely Model Operational Crosstalk}}
Compiler passes rarely model crosstalk which is a major error source \cite{murali2020software, zhao2022quantum}. Quantum operations encounter crosstalk when multiple of them are executed concurrently at the same time. A pass (or multiple passes in conjunction) may arrange gates which minimizes the error according to its (or their) heuristic cost function (such as the cost of SWAPs). However these cost functions don't account for crosstalk and therefore, the final arrangement of gates may suffer from high crosstalk.

CNOT-CNOT crosstalk can be eliminated to a \textit{limited extent} by characterizing crosstalk between CNOT link pairs and serializing the CNOTs over links that form a high-crosstalk pair~\cite{murali2020software}. However, the characterization complexity scales exponential with the system size. Ref \cite{murali2020software} estimates 8 hours of uninterrupted machine time for characterizing crosstalk on a 20-qubit device. This time increases for larger systems, ultimately reaching a point where crosstalk drifts faster than the time it takes to characterize it. To improve scalability, \cite{murali2020software} only characterizes local one-hop apart link pairs. However, our experiments on IBM-Auckland show that there is substantial crosstalk between links multiple hops away. Figure~\ref{fig:combined_motivation_figure}(b) shows that crosstalk causes 1.4x increase in error rate between CNOTs that are 3 hops away, or 1.25x increase between CNOTs 5 hops away. Currently, there are no scalable solutions to estimate this crosstalk, which leads to them not getting accounted for during compilation. 


\vspace{0.05in}
\noindent \textbf{Observation-7:} As compilers lack knowledge about non-local high-crosstalk link pairs, passes schedule concurrent CNOTs on them. For example, in Figure~\ref{fig:combined_motivation_figure}(b), SABRE schedules concurrent SWAPs on links \circled{A} and \circled{B} while relocating distant qubits $Q_0$ and $Q_1$ to minimize circuit depth.

\subsection{\textbf{Existing Fidelity Estimation Tools are Insufficient}}\label{subsec:existing_prediction_tools}
Compilers frequently use the Estimated Success Probability (ESP) \cite{patel2020veritas, tannu2019mitigating, tannu2019ensemble, quetschlich2022compiler, quetschlich2022predicting} metric to assess the probability of successfully executing a schedule analytically using the error-rates of each operation. For successful execution, all gates and measurements must remain error-free and no qubits should decohere. This metric, although easy to compute, ignores phenomena like crosstalk, correlated errors, and device drifts. Also, the error rates from calibration frequently gets outdated due to high device drifts, giving inaccurate ESP estimates. We could also run a noisy simulation of the circuit on a classical computer to estimate the fidelity. This is the most advanced prediction tool available to us. However, it is non-scalable, which makes it a non-feasible solution as circuit sizes grow.  Figure \ref{fig:combined_motivation_figure}(c) shows how the fidelity of four workloads varies with different M-S-R combinations when run on an actual quantum device, noise model or computed using ESP. We perform the device experiments on IBMQ Guadalupe. 

\vspace{0.05in}
\noindent \textbf{Observation-8:} ESP and Noise Model trends have a poor correlation with actual device trends and are not good predictors. 

\section{\textbf{\framework{}: Design}}\label{sec:design}
\begin{figure}[htp]
    \centering
    \includegraphics[width = \columnwidth]{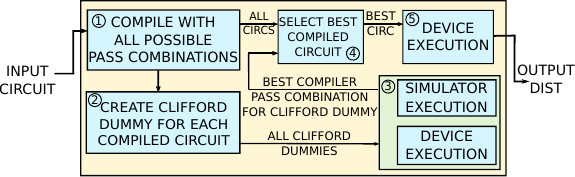}\vspace{-0.1in}
    \caption{Overview of \framework{}.}\vspace{-0.15in}
    \label{fig:optran_overview}
\end{figure}

We propose {\em \framework{}}, a software framework for optimal {\em \underline{Com}piler \underline{Pas}s \underline{S}election} specific to a given program and device at any given time. Figure~\ref{fig:optran_overview} gives an overview of \framework{}. Although this paper focuses on the Qiskit toolchain from IBM, \framework{} is generalizable and can be adapted to equivalent tool-chains from other providers as well. such as Google Cirq, Amazon braket, Rigetti pyQuil, and Quantinuum tket. In addition to the mandatory passes that involve mapping, scheduling, and routing, \framework{} can also be used to decide whether or not to apply optional passes (such as trios or dd). The default implementation of \framework{} includes them. 

\subsection{\textbf{Insights: CNOT Structure Dictates Pass Performance, Leverage Clifford Dummy Circuits}}
\label{subec:clifford_circuits}
\framework{} leverages two key insights. \textit{First}, the performance of a pass combination is mainly dictated by the schedule of the CNOTs generated by them as CNOTs are the dominant source of errors in most quantum programs. \textit{Second}, CNOT is a Clifford gate and thus, transforming a compiled schedule into a Clifford \textit{dummy} captures most of the error characteristics of the schedule. However, unlike the actual schedule, the output of the Clifford dummy can be computed efficiently on classical systems via simulation\cite{aaronson2004improved, knill2008randomized}.

IBM quantum systems support \texttt{ID}, \texttt{X}, \texttt{SX}, \texttt{RZ}, and \texttt{CX} as the basis gates. Therefore, post compiled circuits comprise of only these gates. All these gates are Clifford gates except the \texttt{RZ} gate. \footnote{The RZ gate is often performed virtually. This is achieved by changing the pulse shape of the subsequent gates. So we may deem the RZ gate as a sum of a virtual (software) and non-Clifford component in the next gate.} The \texttt{RZ} gate is a parameterized gate that rotates a qubit around the Z-axis by the angle parameter ($\theta$). If this parameter is in $\{\frac{\pi}{2}, \pi, \frac{3\pi}{2}, 2\pi\}$, the resulting gate is Clifford. For all values in between, \framework{} rounds it to the nearest multiple of $\frac{\pi}{2}$. For parameter values that are close to multiples of $\frac{\pi}{4}$ ($|\theta - \frac{n\pi}{4}| < \delta$), where $\delta$ is a design parameter chosen to be $\frac{\pi}{100}$ and $n \in \{1, 3, 5, 7\}$, \framework{} rounds the parameter to $\frac{(n+1)\pi}{4}$ or $\frac{(n-1)\pi}{4}$ randomly. This process is repeated multiple times with the objective of creating a Clifford circuit with as many peaks in the output distribution as in the original circuit (such as one in Bernstein-Vazirani benchmark). If \framework{} fails to get such a circuit, the one whose number of peaks is closest to the original is chosen. However, in most cases the number of peaks is unknown. For such circuits, we minimize the total number of peaks in the dummy circuit.  While our study mainly involves IBMQ systems, \framework{} can be adapted to other systems by adjusting the basis gate set.

\subsection{\textbf{\framework{}: Description}}\label{subsec:overview}
Next, we describe the key components of \framework{}. 

\vspace{0.05in}
\noindent \circled{1} \textbf{Compile Given Circuit Using All Pass Combinations:} 
This step takes as input the given program and generates multiple executables by using all possible combination of passes. Note that this is essential because each combination yields a different executable with a unique schedule of instructions. If there are $n$ steps in the compiler and the $i^{th}$ step has $a_{i}$ choices, then the total number of circuits would be $\prod_{i = 1}^{n}a_{i}$. For example, if step $i$ corresponds to mapping, then the possible options are \textit{dense (d)}, \textit{noise adaptive (na)} or \textit{sabre (sa)} that gives us $a_{i} = 3$. If step $i$ is an optional physical circuit optimization pass - for example, application of \textit{dynamical decoupling (dd)}, then the two options are the application or non-application of \textit{dd} which gives us $a_{i} = 2$.

\vspace{0.05in}
\noindent \circled{2} \textbf{Create Clifford Dummies For Each Executable:} 
This step create a Clifford Dummy circuit for each of the $\prod_{i = 1}^{n}a_{i}$ compiled circuits. The dummies comprise of only Clifford gates.  Each non-Clifford gate in a compiled circuit is replaced with the ``closest" Clifford gate (as described in Section~\ref{subec:clifford_circuits}). The Clifford gates are left unchanged.

\vspace{0.05in}
\noindent \circled{3} \textbf{Execute Clifford Dummies:} This step simulates the $\prod_{i = 1}^{n}a_{i}$ Clifford dummies on a classical computer and obtains the error-free (\textit{expected}) output. These dummies are also executed on the quantum device. The output fidelity for each dummy is computed by comparing the real machine output with the expected output. Let us denote the pass combination that gives the maximum improvements in fidelity using $P^{*}$.

\vspace{0.05in}
\noindent \circled{4} \textbf{Use Compiled Circuit Corresponding to $P^{*}$:} This step selects the compiled circuit corresponding to the given program and $P^{*}$ pass combination, run this executable on the NISQ machine, and returns the output distribution to the user.


\subsection{\textbf{Why do Clifford Dummy Circuits Work?}}\label{subsec:why_work}

\begin{figure*}[ht!]
    \centering
    \includegraphics[width = 2\columnwidth]{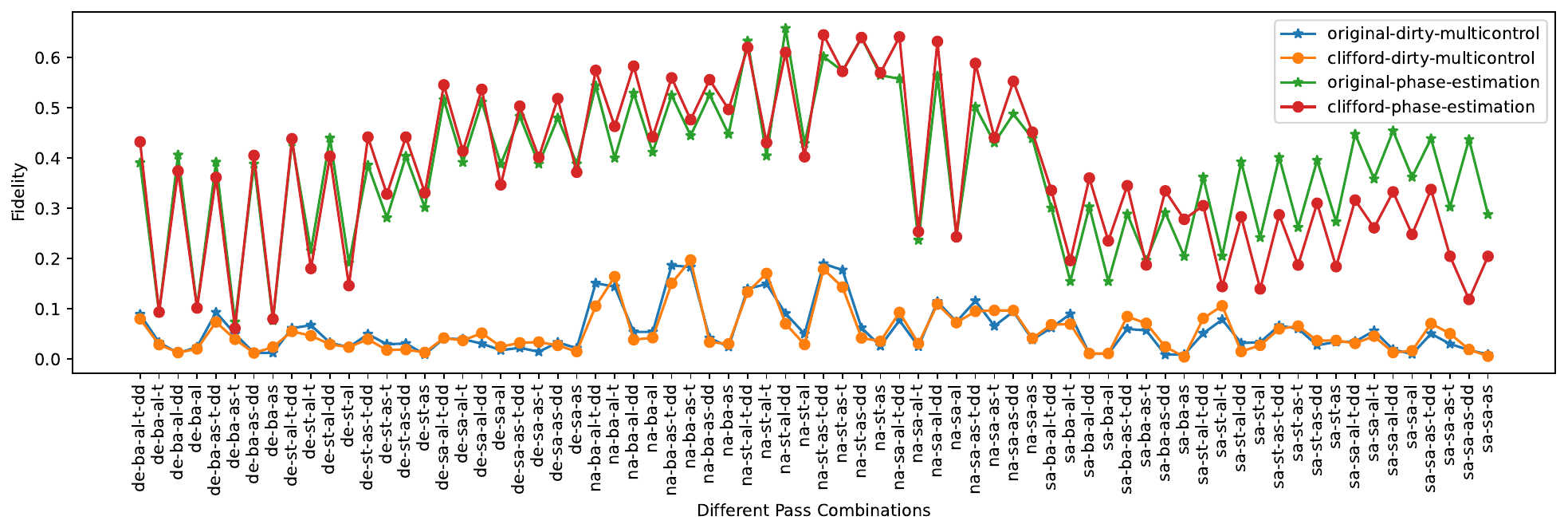}
    \caption{Fidelity of actual circuit and Clifford dummies for phase estimation (pea5) and Multicontrol CNOT (cnxdirty43) benchmarks on IBMQ-Toronto.}\vspace{-0.15in}
    \label{fig:clifford_curve}
\end{figure*}

Structurally, the Clifford Dummies are similar to the compiled versions of the original circuit. CNOT gate errors and crosstalk from CNOTs are dominant sources of errors on NISQ systems~\cite{device_specs}. As CNOT is a Clifford gate, the number, position, and order of execution of CNOTs are the same in the original and dummy circuit. Some passes that we consider have qubit-state-dependent effects. For example, the effect of Dynamical Decoupling on a qubit depends on the state it is in. Thus, we change the parameter of the RZ gate by the smallest amount possible so that the similarity of the qubit state at intermediate positions in the circuit is maximized. We quantify this by checking the correlation between the variation of fidelity with transpiler pass combination in the original and dummy circuits. In most cases, the correlation is extremely strong. Even in cases where it is weak due to a few spurious points, the fidelity of the original and dummy circuit is maximized for the same transpiler pass combination. Figure \ref{fig:clifford_curve} shows the fidelity trend for the original and corresponding dummy circuits for two benchmarks: phase estimation and dirty-multicontrol on IBMQ Toronto. We show the theoretical analysis for the strong correlations in Appendix~\ref{sec:appendix}. 

\section{Efficient \framework{} (\frameworkext{)}}\label{sec:cost}
A key limitation of \framework{} is its poor scalability. The total number of circuit evaluations required for \framework{} scales exponential in the number of transpiler steps. If there are $S$ 
steps in the transpilation pipeline and $a_{i}$ options for the $i^{th}$ step, then \framework{} creates $\Pi_{i = 1}^{S}a_{i}$ Clifford Dummies. As passes often interact with each other in counter-intuitive ways, they cannot be evaluated in isolation. While this motivates the exhaustive search across an exponential number of pass combinations to obtain the best pass combination in \framework{}, it simultaneously limits the scalability. This poses a significant challenge in its adoption. To overcome this challenge, we propose \frameworkext{}, that leverages two key insights, described next, to reduce the overheads of \framework{}.

\subsection{\textbf{Insight-1: Divide-And-Conquer Search}}
\begin{figure}
    \centering
    \includegraphics[width = \columnwidth]{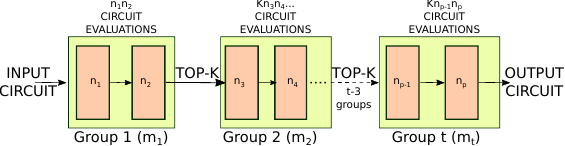}
    \caption{Overview of \frameworkext{}. We split the pipeline into groups ($m_{1}, m_{2}, m_{3}$) and choose the top-k options combinations from each group and feed it into the next, to search for the optimal pass combinations.}
    \label{fig:optran_e_pipeline}
\end{figure}

The isolated or standalone search and the exhaustive search represents two extreme modes of search. The isolated search is scalable but \framework{} cannot account for the interactions between the passes. On the other extreme, using an exhaustive search enables \framework{} to capture all possible interactions between passes but limits the scalability. Ideally, we want to capture the maximum interactions between passes in a scalable manner. To achieve this goal, we propose a divide-and-conquer strategy that achieves a sweet-spot between the isolated and exhaustive search modes. The resultant design, called \frameworkext{}, decomposes the pipeline into $t$ groups, as shown in Figure~\ref{fig:optran_e_pipeline}. \frameworkext{} uses a hyper-parameter $k$ that selects the top $k$ pass combinations from the $i^{th}$ group and sends only those to the $(i+1)^{th}$ group, as starting points of the exhasutive search inside the $(i+1)^{th}$ group. The default implementation uses $k=1$ and $k=3$. We refer to these designs as \frameworkext{}-1 and \frameworkext{}-3.  Each group consists of $m_{j}$ steps where the number of possible pass options available at each step is $n_{i}$, then the number of circuits needed are $\Pi_{i = 1}^{m_{1}}n_{i} + k \cdot \Pi_{i = m_{1}}^{m_{1} + m_{2}}n_{i} +  ... + k \cdot \Pi_{i = \Sigma_{j = 1}^{t-1}m_{j}} ^ {\Sigma_{j = 1}^{t}m_{j}}n_{i}$. Assuming $n_{i} = n \, \forall i$, i.e. each pass stage has an equal number of options to choose from, then the cost reduces to $\mathcal{O}(\frac{p \times n^{t} \times k}{t})$, which\textbf{ scales linear in the number of pipeline stages}, $p$. The default implementation of \frameworkext{} splits the transpiler pipeline into two stages with $m_{1}$ as 3 and $m_{2}$ as 2. The first stage corresponds to mapping, routing, and scheduling. The second stage comprises of additional (optional) optimizations. The  optimizations used in the second stage of the default implementation of \frameworkext{} are Trios and insertion of dynamical decoupling sequences. Note that the choices for these optimizations are to use or not use them. Thus,  the $n_{i}$ values are: $n_{1} = 3, n_{2} = 3, n_{3} = 2, n_{4} = 2, n_{5} = 2$. This reduces the total number of circuits to 18 + $4k$.

\review{}

\subsection{\textbf{Insight-2: Approximate Fidelity Estimation Is Acceptable}}
\label{subsubsec:less_shots}
Quantum circuits are executed for thousands of trials to accurately assess the fidelity and ensure there are no sampling errors in the output distribution. For example, Google uses 25K shots per circuit (with up to 23 qubits) in the QAOA benchmarking studies on the Sycamore machine. However, for identifying the optimal pass combination, the absolute fidelity numbers are not required. Instead, an approximate fidelity that enables us to assess the relative performance of each pass combination against the others is sufficient to identify the optimal combination. Thus, we can tolerate some amount of sampling errors. \frameworkext{} leverages this insight and executes each Clifford dummy circuit for a reduced number of trials or shots. The default \frameworkext{} design allocates a shot budget of 100 shots for every ``peak" in the noise free ideal output distribution of a Clifford dummy circuit. Note that this distribution is already available. This strategy of adapting the number of shots with the number of output peaks minimizes statistical errors in multi-modal distributions. Our evaluations on real systems from IBM show that this shot reduction strategy in \frameworkext{} leads to results at par with the baseline (where we use IBM's default 8K shots for every Clifford Dummy), while drastically reducing the overheads of the search step.


\begin{figure}
    \centering
    \includegraphics[width = \columnwidth]{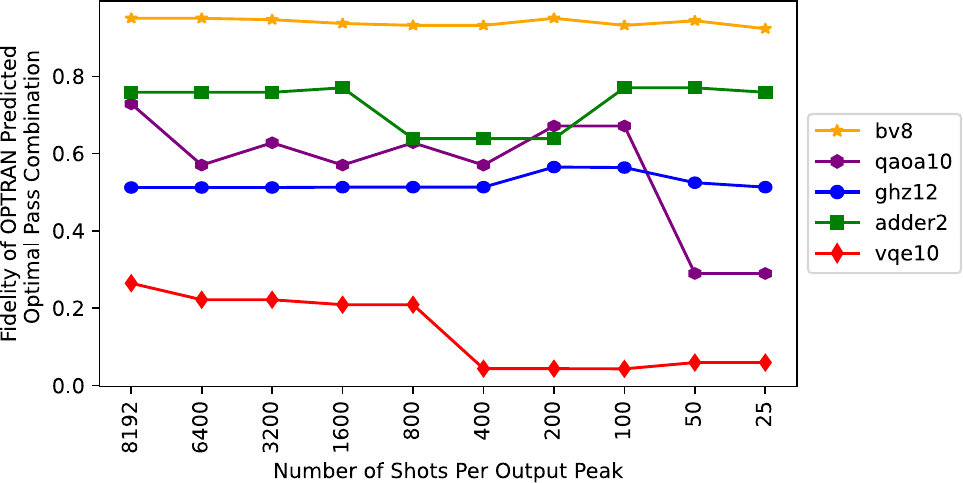}
    \caption{Variation of \framework{} Predicted Optimal Pass Combination Fidelity with ``Shots per Peak" on IBMQ Toronto.}
    \label{fig:variation_with_shots}
\end{figure}

Figure \ref{fig:variation_with_shots} shows how the fidelity corresponding to the optimal pass combination as predicted by \framework{} varies with the number of shots per peak for five workloads on IBMQ Montreal. The trends for \frameworkext{}-3, and \frameworkext{}-1 are similar. Our characterization experiments demonstrate that the fidelities remain fairly stable upto 100 shots. However, for shots fewer than 100, we observe sharper fluctuations indicating the non-negligible impact of statistical errors.

\ignore{

    

\subsection{Overhead}\label{subsec:overhead}

}
\begin{figure*}[t]
    \centering
    \includegraphics[width = \textwidth]{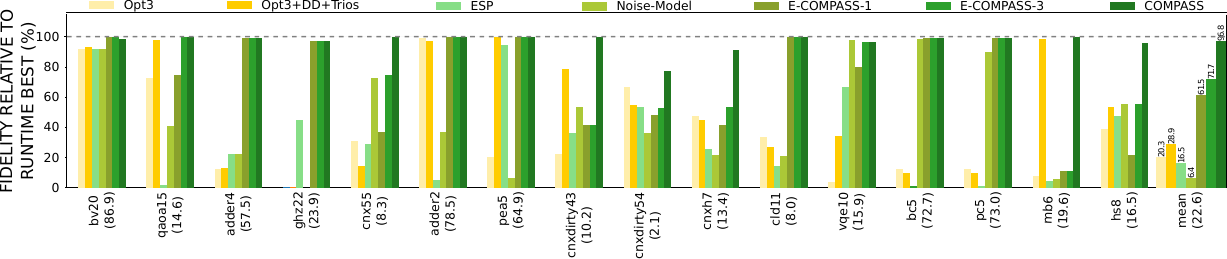}
    \caption{Fidelity relative to Runtime Best on IBM Auckland (Mean fidelities - E-COMPASS-1: 61.5\%, E-COMPASS-3: 71.1\%, COMPASS: 96.8\%).} 
    \label{fig:ibm_auckland_evaluations}
\end{figure*}

\section{\textbf{Methodology}}\label{sec:methodology}



\begin{figure*}[t]
    \centering
    \includegraphics[width = \textwidth]{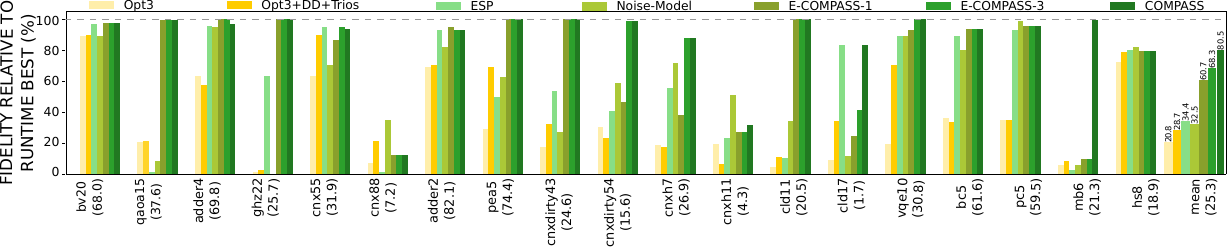}
    \caption{Fidelity relative to Runtime Best on IBM Hanoi (Mean fidelities - E-COMPASS-1: 60.7\%, E-COMPASS-3: 68.3\%, COMPASS: 80.5\%)} 
    \label{fig:ibm_hanoi_evaluations}
\end{figure*}


\subsection{\textbf{Benchmarks}}\label{subsec:benchmarks}
For our evaluations, we use 19 circuits from benchmark suites used in prior works on NISQ compilation~\cite{duckering2021orchestrated, das2021adapt, tomesh2022supermarq, tannu2019not}. The circuit depths and number of qubits are chosen such that each benchmark has an oracle fidelity (maximum possible fidelity across all pass combinations) that is at least $1\%$. If a benchmark has an oracle fidelity below $1\%$ on a particular device, we exclude it for statistical significance. 

\subsection{\textbf{Quantum Hardware}}\label{subsec:hardware}
The final evaluations are conducted on two 27-qubit quantum systems from IBM- IBM Auckland and IBM Hanoi.

\subsection{\textbf{Passes}}\label{subsec:passes}
We consider a comprehensive set of ten representative passes corresponding to different stages in the transpilation pipeline. Thus, we evaluate a total of 72 pass combinations. The passes are:
\begin{enumerate}[leftmargin=0cm,itemindent=.5cm,labelwidth=\itemindent,labelsep=0cm,align=left, itemsep=0.1 cm, listparindent=0.4cm]
    \item Mapping (\texttt{\textbf{M}}) - Dense, Noise Adaptive, Sabre
    \item Scheduling (\texttt{\textbf{S}}) - As Soon As Possible (ASAP), As Late As Possible (ALAP)
    \item Routing (\texttt{\textbf{R}}) - Basic, Stochastic, Sabre
    \item Optimization 1 (\texttt{\textbf{O-1}}) - Trios
    \item Optimization 2 (\texttt{\textbf{O-2}}) - Dynamical Decoupling
\end{enumerate}

As mapping, routing, and scheduling are compulsory steps we choose one pass for each step, giving us 3, 3, and 2 choices at each step respectively. The optimization passes are optional and we have two options for each pass: either apply the optimization or not which gives four options. Thus, we have a total of 3x3x2x2x2 = 72 possible combinations.

\frameworkext{} split the pipeline into two chunks:
\begin{enumerate}[leftmargin=0cm,itemindent=.5cm,labelwidth=\itemindent,labelsep=0cm,align=left, itemsep=0.1 cm, listparindent=0.4cm]
    \item Mapping + Scheduling + Routing (\texttt{\textbf{M-S-R}})
    \item Optimization 1 + Optimization 2 (\texttt{\textbf{O1-O2}})
\end{enumerate}

The default implementation of \frameworkext{} assumes the value of the hyper-parameter $h$ to be 1 and 3 which means only the ``Top 1" or best (for $k=1$) and ``Top 3" (for $k=3$) pass combinations from the first stage (M-S-R) of the transpiler pipeline are combined with the ``Top" combination of the next stage. Note that as the second stage has only four possible options, we only consider the best combination among these. 


\subsection{\textbf{Evaluation Comparisons}}\label{subsec:eval_comparison}
We compare against the different policies for pass set selection enlisted below. 
\begin{enumerate}[leftmargin=0cm,itemindent=.5cm,labelwidth=\itemindent,labelsep=0cm,align=left, itemsep=0.1 cm, listparindent=0.4cm]
    \item \texttt{\textbf{Opt3}}: The $\texttt{optimization level 3}$ flag is the highest possible optimization on IBM-Qiskit (default baseline). 
    \item \texttt{\textbf{Opt3+DD+Trios}}: $\texttt{optimization level 3}$ flag combined with Trios and DD. 
    \item \texttt{\textbf{ESP}}: Pass combination with highest fidelity based on ESP.
    \item \texttt{\textbf{Noise Model}}: Pass combination with highest fidelity based on noisy simulations (noise model created using real device characterization data). Although this method does not scale, it still enables us to have a rigorous comparison. 
    \item \texttt{\textbf{\framework{}}}: Our proposed method that performs an  exhaustive search over all 72 combinations.
    \item \texttt{\textbf{\frameworkext{}-1}}: Our proposed method assuming $h=1$ for the selection of passes from the first stage to the second. 
    \item \texttt{\textbf{\frameworkext{}-3}}: Our proposed method assuming $h=3$ for the selection of passes from the first stage to the second. 
    \item \texttt{\textbf{Runtime Best}}: The combination with the highest fidelity at runtime. This establishes an upper bound on the maximum achievable fidelity on the machine. 
\end{enumerate}

\textit{\textbf{A fidelity closer to that of Runtime Best is desirable. }}

\subsection{\textbf{Evaluation Metrics}}\label{subsec:eval_metrics}


    We use \textbf{Fidelity-Relative-To-Runtime-Best} to assess the performance of our proposed design. We compute fidelities using TVD (Total Variational Distance)~\cite{tvd_wiki}. This metric is derived from prior works~\cite{das2021adapt, das2023imitation, patel2022quest}. We define TVD as the half of the one-norm between the ideal output distribution and the noisy output distribution. $TVD = \frac{1}{2}||P-Q||_{1}$, where $P$ is the ideal distribution co and $Q$ is the noisy distribution. Fidelity-Relative-To-Runtime-Best is defined as $fidelity\_relative\_to\_runtime\_best (M) = \frac{1-TVD(M)}{1-TVD(Runtime Best)}$, where $M$ is the pass selection policy. As ${1-TVD(Runtime Best)}$ is the highest possible fidelity, the resulting number is bounded by 1 (or 100\%).



\section{\textbf{Final Evaluations}}\label{sec:evaluation}

\subsection{Results for Fidelity}
Figure~\ref{fig:ibm_auckland_evaluations} shows the fidelities of the benchmarks for various evaluation modes on IBM Auckland relative to the \textit{Runtime} \textit{Best} policy. The runtime best or maximum fidelity of each circuit corresponds to the $100\%$ mark in the figures and the absolute values are represented are reported along with the benchmark name on the x-axis label. On average, \framework{} achieves $96.86\%$ of the maximum achievable fidelity on IBM Auckland and by up to $100\%$ for many benchmarks. \frameworkext{}-3 and \frameworkext-1 achieve on average $71.77\%$ and $61.52\%$ of the maximum achievable fidelity for a $58.33\%$ and $69.44\%$ reduction in overheads, respectively.

Figure~\ref{fig:ibm_hanoi_evaluations} shows the fidelities on IBM Hanoi relative to the runtime best. On IBM Hanoi, \framework{}, \frameworkext{}-3, and \frameworkext{}-1 achieves $80.51\%$, $68.34\%$, and $60.70\%$ of the maximum achievable fidelity on average. The reduction in cost by \framework{}-3, and \framework{}-1 is  $58.33\%$ and $69.44\%$ respectively for IBM Hanoi as well.

There are some specific cases where \framework{} performs slightly poor compared to other competing policies. However, this is mostly true for circuits with extremely low runtime-best fidelities which results in very minor fluctuations in the absolute values of the fidelities. For example, on IBM Hanoi, the fidelity of the best pass combination reported by the noise model is $56\%$ of the runtime-best fidelity whose absolute value is $2.4\%$. The optimal fidelity reported by \framework{} is $30\%$ of the runtime best which is $1.29\%$. The absolute difference between these numbers is around $1\%$. The distributions for these cases are already too noisy to have meaningful comparisons.

\subsection{Scalability Analysis}
There are four key sources of overheads in \framework{}:
\begin{enumerate}[leftmargin=0cm,itemindent=.5cm,labelwidth=\itemindent,labelsep=0cm,align=left, itemsep=0.1 cm, listparindent=0.4cm]
    \item Compilation Overhead (\texttt{\textbf{Comp}}) - for circuit compilation using all possible pass combinations
    \item Clifford Generation Overhead (\texttt{\textbf{Cliff Gen}}) - for creation of Clifford dummies for each compiled circuit
    \item Simulation Overhead (\texttt{\textbf{Cliff Sim}}) - for executing the dummies on classical simulator
    \item Quantum Execution Overhead (\texttt{\textbf{QED}}) - for executing the dummies on the real quantum hardware
\end{enumerate}

The overhead for all these steps is documented in Table \ref{tab:optrane1_overhead_hanoi}. For steps (1)-(3), we provide the wall clock time because these steps are executed on a classical computer. The simulation step (3) is performed using Qiskit's stabilizer simulator. For (4), we provide the total number of shots needed to execute all the Clifford dummies, as a fraction of the number of shots used while executing the baseline (10,000).

We observe that the overheads are fairly nominal compared to the potential fidelity improvement from \framework{}. The maximum net classical overhead is in the order of a hundred seconds. For most programs, it is an order of magnitude lesser than the maximum classical overhead. The quantum execution overhead for \framework{}, and \frameworkext{} of all benchmarks is reported in Table~\ref{tab:optrane1_overhead_hanoi} as well. Note that we use approximate fidelity estimation for baseline \framework{} as well in order to reduce shot overhead to enable a fair comparison. For most cases, it is less than $0.5\times$ the baseline. The maximum quantum overhead is $1.36\times$. This is a substantial reduction from 72x and 22x overhead that \framework{} and \frameworkext{}-1 introduce respectively without the approximate estimation. The Clifford simulation time is especially small, which is expected.

\begin{table*}[]
    \centering
    \renewcommand{\arraystretch}{1.2}
    \setlength{\tabcolsep}{10pt}
    \begin{tabular}{|c|c|c|c|c|c|c|c|c|c|c|}
    \hline
        \multirow{2}{*}{Circuit} & \multirow{2}{*}{Qubits} & \multirow{2}{*}{Depth} & \multirow{2}{*}{Gates} & Non Clifford & Comp & Cliff & Cliff & QED & QED & QED \\
         & & & &  Gates & [s] & Gen [s] & Sim [s] & (C) & (CE-3) & (CE-1)\\
         
         \hline
         bv20 & 20 & 2-6 & 28-32 & 0 & 2.03 & 4.44 & 1.18 & 0.72x & 0.30x & 0.22x\\
         \hline
         qaoa15 & 15 & 52-202 & 202-875 & 0-2 & 39.71 & 2.93 & 4.55 & 0.72x & 0.30x & 0.22x\\
         \hline
         adder4 & 4 & 45-141 & 136-437 & 14-18 & 11.68 & 2.21 & 3.29 & 0.72x & 0.30x & 0.22x\\
         \hline
         ghz22 & 22 & 45-234 & 142-654 & 0 & 14.39 & 16.40 & 3.20 & 1.44x & 0.60x & 0.44x\\
         \hline
         cnx55 & 10 & 72-259 & 262-828 & 45-65 & 20.37 & 69.54 & 5.48 & 0.88x & 0.37x &0.27x\\
         \hline
         cnx88 & 16 & 92-321 & 446-1546 & 81-99 & 36.34 & 181.76 & 7.91 & 2.35x & 0.98x &0.72x\\
         \hline
         adder2 & 4 & 37-96 & 116-336 & 14-19 & 6.99 & 3.09 & 2.62 & 0.72x & 0.30x &0.22x\\
         \hline
         pea5 & 5 & 62-102 & 162-367 & 21-27 & 15.78 & 11.27 & 5.23 & 0.72x & 0.30x &0.22x\\
         \hline
         cnxdirty43 & 8 & 102-327 & 256-922 & 50-67 & 22.61 & 38.93 & 5.82 & 0.79x & 0.32x &0.24x\\
         \hline
         cnxdirty54 & 10 & 172-488 & 385-1429 & 74-94 & 32.23 & 135.38 & 7.82 & 1.14x & 0.48x &0.35x\\
         \hline
         cnxh7 & 7 & 119-315 & 278-849 & 50-62 & 22.18 & 41.32 & 5.82 & 0.79x & 0.32x &0.24x\\
         \hline
         cnxh11 & 11 & 227-581 & 496-1708 & 102-128 & 44.33 & 176.24 & 11.75 & 1.51x & 0.62x &0.46x\\
         \hline
         cld11 & 11 & 74-344 & 316-1153 & 55-76 & 25.84 & 136.88 & 6.48 & 1.41x & 0.58x & 0.43x\\
         \hline
         cld17 & 17 & 126-475 & 572-1850 & 95-130 & 44.45 & 252.13 & 9.38 & 4.45x & 1.85x & 1.36x\\
         \hline
         vqe10 & 10 & 23-145 & 161-523 & 30-47 & 10.01 & 2.02 & 3.09 & 0.72x & 0.30x & 0.22x\\
         \hline
         bc5 & 5 & 18-72 & 96-280 & 0 & 3.76 & 10.12 & 2.09 & 0.72x & 0.30x & 0.22x\\
         \hline
         pc5 & 5 & 27-79 & 114-324 & 2-4 & 7.02 & 20.77 & 2.28 & 0.72x & 0.30x & 0.22x\\
         \hline
         mb6 & 6 & 112-230 & 244-678 & 2-11 & 22.34 & 2.008 & 4.28 & 0.72x & 0.30x & 0.22x\\
         \hline
         hs8 & 8 & 33-119 & 139-457 & 10-25 & 9.08 & 85.17 & 7.93 & 1.60x & 0.67x & 0.49x\\
         \hline
    \end{tabular} \vspace{0.1in}
    \caption{\vspace{0.05in} {Benchmark specifications and overhead for \framework{} (C), \frameworkext{}-3 (CE-3), and \frameworkext{}-1 (CE-1) on IBM Hanoi. The Compilation, Clifford Generation, and Simulation Overheads are reported in seconds. The Device Execution Overhead is reported as the number of extra shots needed to run \framework{} relative to the baseline number of shots (10,000).}}
    \label{tab:optrane1_overhead_hanoi}\vspace{-0.25in}
\end{table*}

\framework{} is scalable even for benchmarks with 100 qubits and 200,000+ gates NISQ circuits. Note that we can \textit{\textbf{run steps (1)-(3) in parallel}} on different machines. The time reported in Table~\ref{tab:scalability_overhead} assuming this parallel implementation of \framework{}.

\begin{table}[htp]
    \centering
    \renewcommand{\arraystretch}{1.2}
    \setlength{\tabcolsep}{6pt}
    \begin{tabular}{|c|c|c|c|c|c|}
    \hline
         \multirow{2}{*}{Circuit} &  \multirow{2}{*}{Qubits} & Gates & \multirow{2}{*}{Comp (s)} & Cliff & Cliff \\
         & & (1k=1000)& & Gen (s) & Sim (s)\\
         \hline
         qaoa40 & 40 & 10.5k-27.7k & 27.6 & 0.17 & 1.85\\
         \hline
         qaoa70 & 70 & 41.8k-91.6k & 110.99 & 0.41 & 4.11\\
         \hline
         qaoa100 & 100 & 92.5k-202.4k& 242.63 & 0.72 & 7.1\\
         \hline
         cnx40 & 40 & 3.3k-12.7k & 6.59 & 2.94 & 2.73 \\
         \hline
         cnx70 & 70 & 7.1k-33.4k & 12.23 & 6.46 & 6.60\\
         \hline
         cnx100 & 100 & 8.9k-55.3k & 20.88 & 11.85 & 11.23 \\
         \hline
         cnxh39 & 39 & 6.6k-17.4k & 10.99  & 3.42 & 2.63 \\
         \hline
         cnxh99 & 99 & 19.3k-83.7k & 29.84 & 12.78 & 12.50 \\
         \hline
         cld39 & 39 & 2.84k-14.4k & 5.91 & 2.75 & 2.29 \\
         \hline
         cld69 & 69 & 6.4k-30.2k & 11.03 & 6.29 & 6.26 \\
         \hline
         cld99 & 99 & 9.15k-51.4k & 20.07 & 11.13 & 12.34\\
         \hline
    \end{tabular}  \vspace{0.1in}
    \caption{ \vspace{0.05in}{Classical Overhead for \framework{} for larger NISQ circuits}}
    \label{tab:scalability_overhead} \vspace{-0.3in}
\end{table}

\subsection{Performance Of Clifford Dummies}
\framework{} predicts the best pass combinations well because of the high correlation in the fidelity trends between the original circuit and the Clifford circuits. Table \ref{tab:hanoi_correlation} lists the correlation values when evaluated on IBM Hanoi for a few benchmarks. We also prove that this correlation will be high assuming a global depolarizing noise channel in Appendix~\ref{sec:appendix}.

\begin{table}[htp]
    \centering
    \renewcommand{\arraystretch}{1.2}
    \setlength{\tabcolsep}{6pt}
    \begin{tabular}{|c|c|}
    \hline
         Circuit &  Correlation ($\%$) \\
         \hline
         bv20 & 97.29\\ 
         \hline
         qaoa15 & 99.93\\ 
         \hline
         adder4 & 99.15\\ 
         \hline
         ghz22 & 99.88\\ 
         \hline
         cnx88 & 69.98\\ 
         \hline
         cnxdirty54 & 78.16\\ 
         \hline
         cld17 & 79.91 \\
         \hline
         hs8 & 78.26\\ 
         \hline
         vqe10 & 85.88\\ 
         \hline
    \end{tabular}  \vspace{0.1in}
    \caption{ \vspace{0.05in} Correlation data for IBM Hanoi}
    \label{tab:hanoi_correlation}\vspace{-0.35in}
\end{table} 

\subsection{Scalability/Adaptability to Other Methods}
\label{subec:adapting}
Software error mitigation is an active area of research for NISQ applications with newer methods being frequently developed. To assess the feasibility of expanding \framework{} to other methods, we perform a \textit{case study} using ensemble-based error mitigation schemes~\cite{tannu2019ensemble,patel2020veritas}. These methods create ensembles of different \textit{mappings} and combine the output distributions to alleviate correlated errors. For a fair assessment, we extend this idea beyond just mapping and include all pass combinations. We consider a \textit{tuple} of the pass options used at each stage of the transpiler pipeline as an instance and create an ensemble of these instances. Also, instead of using the device calibration data to rank the instances, as proposed originally~\cite{tannu2019ensemble, patel2020veritas} to rank mappings, we use the Clifford Dummy score because we have already established that dummies offer better estimates than ESP. 

\begin{figure}[h]
    \centering
    \includegraphics[width = 0.9\columnwidth]{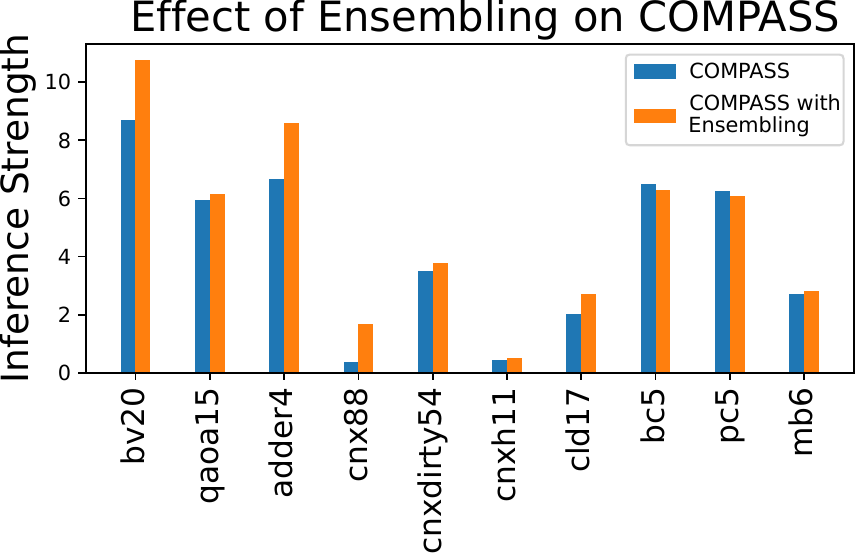}\vspace{-0.1in}
    \caption{Inference Strength (IST) \cite{tannu2019ensemble} of workloads increases when using an ensemble of the ``top-k" \framework{} predictions instead of the best one only.} \vspace{-0.1in}
    \label{fig:optran_plus_edm}
\end{figure}

We evaluate the combined design that extends \framework{} to include ensemble techniques. Note that the objective of ensembles is not to increase the absolute fidelity but to improve it relative to the most frequently occurring incorrect output. This is called ``\textit{Inference Strength (IST)}" \cite{tannu2019ensemble}. We use the same metric to enable a fair comparison. An IST greater than 1 and higher IST is desirable. Figure~\ref{fig:optran_plus_edm} shows the IST for \framework{} and the enhanced version of \framework{}.

\section{\textbf{Related Work}}\label{sec:discussion}

\noindent \textbf{Clifford dummies in NISQ:} ADAPT and Imitation Game~\cite{das2021adapt, das2023imitation} employ Clifford circuits to identify the optimal usage of dynamical decoupling sequences and native gate combination for gate decomposition respectively. These passes are additional optimizations beyond the mapping, routing, and scheduling steps, similar to Trios and DD used in \framework{}. \framework{} is orthogonal to the above techniques as it employs Clifford circuits to identify the optimal subset of optimizations during compilations. ADAPT and Imitation Game can be included as additional stages in \framework{}  (similar to the integration of the ensemble techniques in Section~\ref{subec:adapting}), we reserve it for future work. Moreover, the Clifford dummies used in \framework{} does not employ any non-Clifford gates unlike ADAPT and Imitation Game. Unlike DD insertion or native gate selection which tends to depend on the state of the qubits, the performance of compiler passes is heavily dictated by the structure of the CNOTs and therefore, using only Clifford gates suffice. \framework{} also uses additional overhead reduction techniques such as approximate fidelity estimation. Therefore, \framework{} is more scalable. Other than this, Clifford circuits have been used as stabilizers in quantum error correction \cite{aaronson2004improved, fowler2012surface}, error mitigation~\cite{ravi2022boosting}, classical simulation~\cite{cafqa}, quantum networks \cite{Veitch2014} and teleportation \cite{yoshida2019disentangling}.

\ignore{
in a philosophically similar way as COMPASS - as a tool to score/rank multiple different Dynamical Decoupling (DD) sequences, and Native entangling gates respectively. However, the novelty of COMPASS lies in the problem-solution fit for Clifford circuits. Both the task of finding optimal entangling gates and DD sequences are heavily dependant on qubit state \cite{pokharel2018demonstration, das2021adapt} which gets altered during conversion to a Clifford Dummy while most compiler passes make decisions based on the CNOT skeleton of the circuit which remains unchanged since CNOT is a Clifford gate. As a result, \cite{das2021adapt, das2023imitation} have to insert non-Clifford gates in their ``Clifford circuit" to get a good correlation, which makes the technique non-scalable.
}

\noindent \textbf{Optimal pass selection in NISQ compilation:} Selecting the optimal pass set involves a \textit{search} that \textit{ranks} all candidates based on an estimated fidelity cost. In \framework{}, this  cost is determined from the performance of the Clifford dummies. Recent works focus on the search problem using machine learning (ML) and reinforcement learning (RL)~\cite{quetschlich2022predicting, quetschlich2022compiler}. However, their performance is limited by their ESP-based fidelity estimation which gives sub-par results even with an exhaustive search. On the other hand, \framework{} uses Clifford dummies that provide much more accurate fidelity estimates specific to a program and machine. Note that the capabilities of~\cite{quetschlich2022compiler, quetschlich2022predicting} is also severely limited due to an even narrower search space, guided by the ML model or the RL agent, and hence the solution is guaranteed to be only as good as that predicted by ESP using an exhaustive search. \framework{} does not have such drawbacks as it does not use ESP at all.

\section{\textbf{Conclusion}}\label{sec:conclusion}
NISQ compilers play a crucial role in improving application fidelity by leveraging software optimizations or passes to generate highly optimized machine code during program translation. 
This paper proposes \framework{}, a fully automated software framework for identifying the optimal set of compiler passes specific to a program and a given machine. \framework{} uses Clifford dummy circuits (that mimic the compiled program but has a known output) to evaluate the performance of each pass combination and identifies the one that maximizes fidelity. Our experiments on IBM machines show that COMPASS improves fidelity by $88.68\%$ of the maximum possible limit over the baseline used by IBM Qiskit. We also propose \frameworkext{} that reduces the search overheads of \framework{} by using a divide-and-conquer search and approximate fidelity estimation methods. \frameworkext{} improves the fidelity by $70.055\%$ of the maximum achievable limit over the baseline with $58.33\%$ lower overheads compared to \framework{}. 
\section{\textbf{Appendix}}\label{sec:appendix}
Here we provide a theoretical argument supporting the use of Clifford circuits for evaluating compiler passes. Let $c_{1}, c_{2}, ..., c_{n}$ be the $n$ compiler pass combinations, let $\rho$ be the output state of the original input circuit, let $\epsilon_{i}$ be an error channel solely determined by the circuit structure obtained by pass combination $c_{i}$, and let $U_{i}=e^{-i\delta H_{i}}$ (with $|H|=1$) be the unitary that rotates the state $\rho$ to its nearest Clifford when compiled by $c_{i}$. Then define $f_i = Tr[\rho\epsilon_{i}(\rho)]$ and $f_i^{\prime} = Tr[U_i \rho U_i^{\dagger}\epsilon_{i}(U_i\rho U_i^{\dagger})]$,
which fidelity of $\rho$ and $U_i\rho U_i^{\dagger}$, respectively, under the error $\epsilon_i$.

Now, to first order in $\delta$ we have $U_{i} \approx I - i\delta H_{i}$, and 
\\ $U_{i}\rho U_{i}^{\dagger} = (I - i\delta H_{i})\rho(I + i\delta H_{i}) \approx \rho + i\delta[\rho, H_{i}]$, 

and the difference in fidelities to first order is given by \\
$|f_{i} - f_{i}^{'}| = |Tr[\rho\epsilon_{i}(\rho)] - Tr[U_{i}\rho U_{i}^{\dagger}\epsilon_{i}([U_{i}\rho U_{i}^{\dagger})]|  \\
\approx |Tr[\rho\epsilon_{i}(\rho) - (\rho\epsilon_{i}(\rho) + i\delta(\rho\epsilon_{i}([\rho, H_{i}]) + [\rho, H_{i}]\epsilon_{i}(\rho)))]| \\
 = \delta|Tr[i([\rho, H_{i}](\epsilon^{\#}_{i}(\rho) + \epsilon_{i}(\rho)))]|$,
 
 where $\epsilon^{\#}$ denotes the dual map of $\epsilon$.

Let us now consider self-dual error maps, where $\epsilon^{\#}=\epsilon$. Then we can decompose the error map into a component that acts as $\epsilon(\rho)=\rho$ with weight $1-p$ and a non-trivial component $\tilde{\epsilon}$ with weight $p$. (For a non-self-dual map we can perform this decomposition on the map and its dual separately). Then, we have 

$|2\delta((1-p)Tr[i[\rho, H_{i}]\rho] + pTr[i\tilde{\epsilon_i}(\rho)])|$. \\
Then, $Tr[i[\rho, H]\rho] = Tr[i\rho H \rho - iH\rho \rho] = iTr[\rho H \rho] - iTr[H \rho \rho] = 0$ by the cyclic property of trace, so we have $2 \delta p Tr[i\tilde{\epsilon_i}(\rho)]$. This shows in a compact form that as $\delta \rightarrow 0$, (i.e. Clifford approximation is not far from original circuit), and as $p \rightarrow 0$ (the error map is not too strong), $|f_{i} - f_{i}^{'}| \rightarrow 0$, indicating that there is a perfect correlation between the original state $\rho$ and the Clifford circuit closest to $\rho$ compiled under $c_i$. When $p$ and $\delta$ do not go to zero, the efficacy of our approach then depends to first order in $\delta$ on how this difference in fidelities of one compiler $c_i$ scales with respect to other compilers, which depends non-trivially on the underlying problem, compiler passes, and noise model.

\section{Acknowledgements}\label{sec:appendix}
This work is funded in part by EPiQC, an NSF Expedition in Computing, under award CCF-1730449; in part by STAQ under award NSF Phy-1818914/232580; in part by the US Department of Energy Office of Advanced Scientific Computing Research, Accelerated 
Research for Quantum Computing Program; and in part by the NSF Quantum Leap Challenge Institute for Hybrid Quantum Architectures and Networks (NSF Award 2016136), in part based upon work supported by the U.S. Department of Energy, Office of Science, National Quantum 
Information Science Research Centers, and in part by the Army Research Office under Grant Number W911NF-23-1-0077. The views and conclusions contained in this document are those of the authors and should not be interpreted as representing the official policies, either expressed or implied, of the U.S. Government. The U.S. Government is authorized to reproduce and distribute reprints for Government purposes notwithstanding any copyright notation herein. This research used resources of the Oak Ridge Leadership Computing Facility, which is a DOE Office of Science User Facility supported under Contract DE-AC05-00OR22725. FTC is the Chief Scientist for Quantum Software at Infleqtion and an advisor to Quantum Circuits, Inc.

\bibliography{refs}

\end{document}